\documentclass[12pt]{iopart}
\usepackage{graphicx}
\usepackage{amsfonts}
\usepackage{graphicx}
\usepackage{graphics}
\usepackage{iopams}
\usepackage{subfig}
\usepackage{dsfont}
\usepackage[usenames]{color}

\def\bea{\begin{eqnarray}}
\def\eea{\end{eqnarray}}

\begin{document}

\title{Rotational symmetry breaking in baby Skyrme models}

\author{I Hen and M Karliner}
\address{School of Physics and Astronomy,
Raymond and Beverly Sackler Faculty of Exact Sciences,
Tel-Aviv University, Tel-Aviv 69978,
Israel.}
\ead{\mailto{itayhe@post.tau.ac.il}, \mailto{marek@neutron.tau.ac.il}} 
\begin{abstract}
We consider multisolitons with charges $1 \leq B \leq 5$ in the baby Skyrme model for the one--parametric 
family  of potentials $U=\mu^2 (1-\phi_3)^s$ with $0<s \leq 4$.
This class of potentials
is a generalization of the `old' ($s=1$) and `holomorphic' ($s=4$) baby Skyrme models.
We find that for charge one, stable solutions exist 
for every value of $s$ and they are rotationally-symmetric. 
For higher charges, stable solutions exist only below $s \approx 2$.
In the charge--two sector the stable solutions are always 
rotationally--symmetric and ring--like. For charge three and above,
rotational symmetry is exhibited only in the small $s$ region; 
above a certain critical value of $s$, this symmetry is broken 
and a strong repulsion between the constituent one--Skyrmions becomes apparent.
We also compute the spatial energy distributions of these solutions.
\end{abstract}

\pacs{0350, 1110}
\submitto{Nonlinearity}
\maketitle

\section{Introduction}

The Skyrme model \cite{Skyrme1,Skyrme2} is an SU(2)--valued non-linear theory for pions in (3+1) dimensions
with topological soliton solutions called Skyrmions. 
Apart from a kinetic term, the Lagrangian of the model contains 
a `Skyrme' term which is of the fourth order in derivatives, 
and is used to introduce scale to the model \cite{3DSkyrme}. 
The existence of stable solutions in the Skyrme model is a consequence of the nontrivial
topology of the mapping $\mathcal{M}$ of the physical space
into the field space  at a given time,
$\quad \mathcal{M}:  S^3 \to SU(2) \cong S^3,\quad$ where 
the physical space $\mathbb{R}^3$  is compactified to $S^3$
by requiring the spatial infinity to be equivalent in
each direction.
The topology which stems from this one-point 
compactification allows the classification of maps into equivalence classes,
each of which has a unique conserved quantity called the topological charge.
\par
The Skyrme model has an analogue in (2+1) dimensions known as the baby Skyrme model,
also admitting stable field configurations of a solitonic nature \cite{Old1}.
Due to its lower dimension, it serves as a simplification of the original model,
but nonetheless it has a physical significance on its own,
having several potential applications in condensed-matter physics \cite{Condensed}.
The target manifold in the baby model is described by a three-dimensional vector 
$\bphi$  with the constraint $\bphi \cdot \bphi=1$. 
In analogy with the $(3+1)$D case,
the domain of this model $\mathbb{R}^2$ is compactified to $S^2$
yielding the topology required
for the classification of its field configurations into classes with conserved topological charges.
The Lagrangian density of the baby Skyrme model is given by:
\bea \label{eq:BabyLag}
\fl \mathcal{L}=\frac1{2} \partial_{\mu} \bphi \cdot \partial^{\mu} \bphi
- \frac{\kappa^2}{2}\big[(\partial_{\mu} \bphi \cdot \partial^{\mu} \bphi)^2 -
(\partial_{\mu} \bphi \cdot \partial_{\nu} \bphi) 
\cdot (\partial^{\mu} \bphi \cdot \partial^{\nu} \bphi)
\big]
-U(\phi_3)
\eea
and consists analogously of a kinetic, the Skyrme and a potential term. 
\par
While in (3+1) dimensions the latter term is optional \cite{TopoSol},
its presence in the (2+1)D model is necessary
for the stability of the solutions.
Aside from the requirement that the potential vanishes at infinity for a given vacuum field
value (normally taken to be $\bphi^{(0)}=(0,0,1)$), 
its exact form is arbitrary and gives rise to a rich family
of possible baby-Skyrme models. Three specific
potentials have been studied in  detail in the literature.
The simplest is the holomorphic model with
$U(\phi_3)=\mu^2 (1-\phi_3)^4$ \cite{Holo1,Holo2,Holo3}.
It is known to have a stable solution only in the charge--one sector. 
The model with the potential
$U(\phi_3)=\mu^2 (1-\phi_3)$ (commonly referred to as the `old' model) 
has also been extensively studied.
This potential gives rise to very structured non-rotationally-symmetric 
multi-Skyrmions \cite{Old1,Old2}.
Another model with $U(\phi_3)=\mu^2 (1-\phi_3^2)$ produces ring-like multi-Skyrmions \cite{Weidig}.
Other double--vacuum potentials which give rise to other types of solutions
have been studied in \cite{babyClass}. 
\par
The brief review given above indicates 
that the form of the potential term
has a decisive effect on the properties of the minimal energy configurations
of the model.
To gain better insight into this matter,
in this paper we study the multisolitons of the baby Skyrme model for the one--parametric 
family of potentials $U=\mu^2 (1-\phi_3)^s$ with $0<s \leq 4$ which generalizes
previously studied cases. Setting $s=1$ yields the `old` model,
whereas $s=4$ corresponds to the holomorphic model.
Although at this time the physical motivation for these
generalizations is not yet fully clear, we show that it nonetheless
yields interesting results. 
The value of the parameter $s$ 
has dramatic effects on the static solutions
of the model, both quantitatively and qualitatively, in the sense
that it can be viewed as a `control' parameter responsible
for the repulsion or attraction between Skyrmions.

\section{\label{sec:method} Method}
In the present work, we find the static multisolitons of the baby Skyrme 
model whose Lagrangian density is given by:
\bea \label{eq:BabyLagFamily}
\fl \mathcal{L}=\frac1{2} \partial_{\mu} \bphi \cdot \partial^{\mu} \bphi
- \frac{\kappa^2}{2}\left( (\partial_{\mu} \bphi \cdot \partial^{\mu} \bphi)^2 -
(\partial_{\mu} \bphi \cdot \partial_{\nu} \bphi) 
\cdot (\partial^{\mu} \bphi \cdot \partial^{\nu} \bphi)
\right)
-\mu^2 (1-\phi_3)^s \;.
\eea
The model contains three free parameters, namely $\kappa, \mu$ and $s$.
Since either $\kappa$ or $\mu$ may be scaled away,
the parameter space of this model is in fact only two dimensional.
Our main goal in this paper is to study the effects of these parameters
on the static solutions of the model within each topological sector.  \par
The multi--Skyrmions of our model
are those field configurations which minimize the static energy
functional within each topological sector. 
In polar coordinates the energy functional is given by
\bea \label{eq:O3energy}
\fl E=\int r \, \rmd r \, \rmd \theta \left(
\frac1{2} (\partial_r \bphi \cdot \partial_r \bphi 
+ \frac1{r^2}\partial_{\theta} \bphi \cdot \partial_{\theta} \bphi)
+\frac{\kappa^2}{2} \frac{(\partial_r \bphi \times \partial_{\theta} \bphi)^2}{r^2}
+\mu^2 (1-\phi_3)^s
\right) \;.
\eea
\par
The Euler--Lagrange equations
derived from the energy functional (\ref{eq:O3energy})
are non--linear \textit{PDE}'s, so in most cases one must resort to 
numerical techniques in order to solve them. In our 
approach, the minimal energy configuration of a baby Skyrmion of charge B 
and a given set of values $\mu, \kappa, s$
is found by a full--field relaxation method, described in more detail below.\par
One of our objectives is the study of spontaneous breaking of rotational symmetry 
by the static solutions.
To avoid a possible bias from the use of a rectangular grid in
cases in which the rotational symmetry of the solution is broken,
a polar grid has been used -- $120$ grid points for the radial coordinate 
and $120$ points for the angular coordinate.
For numerical convenience, the physical $\mathbb{R}^2$ plane is further compactified into the unit circle
by transforming to a truncated radial coordinate
\bea \label{eq:truncR}
\tilde{r}=\frac{2}{\pi} \tan^{-1}(r/\alpha)
\eea
with $\alpha=1.2$. The relaxation process begins by initializing the field triplet $\bphi$
to a rotationally--symmetric configuration  
\bea \label{eq:initConfig}
\bphi_{\textrm{initial}}=(\sin f(r) \cos B \theta,\sin f(r) \sin B \theta,\cos f(r)) \quad,
\eea
where $B$ is the topological charge of the baby--Skyrmion in question
and the profile function $f(r)$ is set to $f(r)=\pi \exp(-r)$.
\par
The energy of the baby--Skyrmion is then minimized by repeating the following steps:
a point $(\tilde{r}_m,\theta_n)$ on the grid is chosen at random, along 
with one of the three components of the field $\bphi(\tilde{r}_m,\theta_n)$.
The chosen component is then shifted by a value $\delta_{\phi}$ chosen uniformly from the segment $[-\Delta_{\phi},\Delta_{\phi}]$
where $\Delta_{\phi}=0.1$ initially. The field triplet is then normalized
and the change in energy is calculated.
If the energy decreases, the modification of the field is accepted
and otherwise it is discarded.
The procedure is repeated while the value of 
$\Delta_{\phi}$ is gradually decreased throughout
the procedure. This is done until no further decrease in energy is observed.\par
One undesired feature of this minimization scheme
is that it can get stuck at a local minimum.
This problem can be resolved by using the ``simulated annealing'' algorithm \cite{SA1,SA2},
which in fact has been successfully implemented before,
in obtaining the minimal energy configurations of 
static two and three dimensional Skyrmions \cite{SkyrmeSA}. 
The algorithm comprises of repeated applications of a Metropolis algorithm 
with a gradually decreasing temperature,
based on the fact that when a physical system is slowly cooled down,
reaching thermal equilibrium at each temperature, it will end up in its ground state. 
This algorithm, however, is much more expensive in terms of computer time. 
We therefore employ it only on a representative sample of the parameter space,
just as a check on our results, which correspond to a Metropolis algorithm of zero temperature.
\par
As a further verification, we set up the minimization scheme using
different initial configurations and grids of different sizes ($60 \times 60$ and $90 \times 90$)
for several $\kappa,\mu$ and $s$ values. 
This was done to make sure that the final configurations are independent of the
discretization and cooling scheme. Accuracy  was also verified by checking conservation
of the topological charge $B$ throughout the minimization process, yielding
$\displaystyle{\Big| \frac{B_{\textrm{{\scriptsize observed}}}}{B} - 1 \Big|< 10^{-5}}$.

\section{Results}
Using the minimization procedure presented in the previous section, 
we have found the static solutions of the model for $1 \leq B \leq 5$.
Since the parameter space of the model is effectively two dimensional (as discussed in section \ref{sec:method}),
without loss of generality 
we fixed the potential strength at $\mu^2=0.1$ throughout,
and scanned the $s$--$\kappa$ parameter space in
the region $0<s \leq 4$ and $0.01 \leq \kappa^2 \leq 1$.

\subsection{Charge--one Skyrmions}
\begin{figure}[htp!]
\includegraphics[angle=0,scale=1,width=1.1\textwidth]{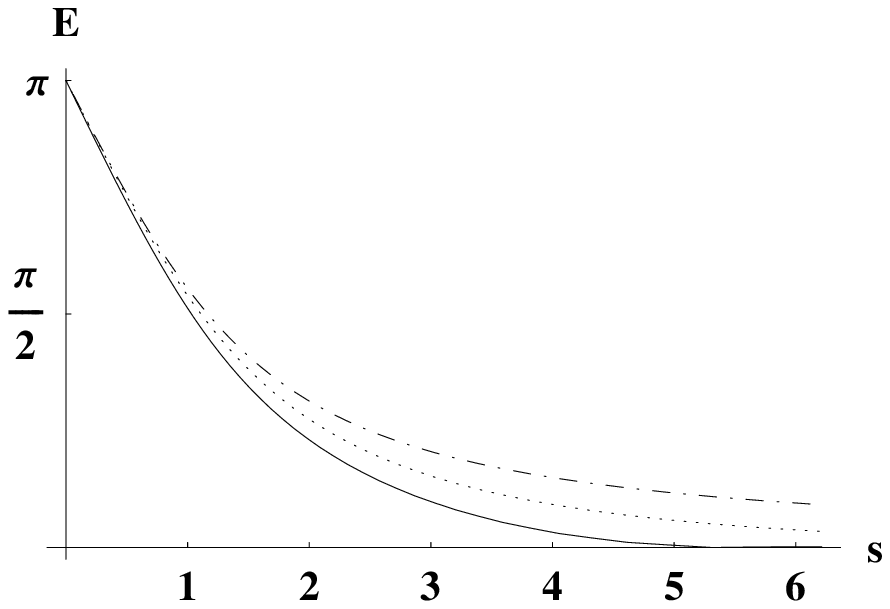}
\caption{Profile functions of the $B=1$ Skyrmion
for $s=0.5$ ($\full$), $s=1$ ($\dotted$) and $s=2$ ($\chain$). Here $\kappa$ is fixed at $\kappa^2=0.25$.}
\label{profB1}
\end{figure}

In the charge--one sector, stable rotationally-symmetric solutions 
were found for all tested values of $s$ and $\kappa$. 
Figure \ref{profB1} shows the obtained profile functions of the $B=1$ solution for 
different values of $s$ with $\kappa$ fixed at $\kappa^2=0.25$. 
\par
\begin{center}
\begin{figure}[hbp!]
\subfloat[$\kappa^2=0.05$]{
\label{s123a} 
\includegraphics[angle=0,scale=1,width=0.60\textwidth]{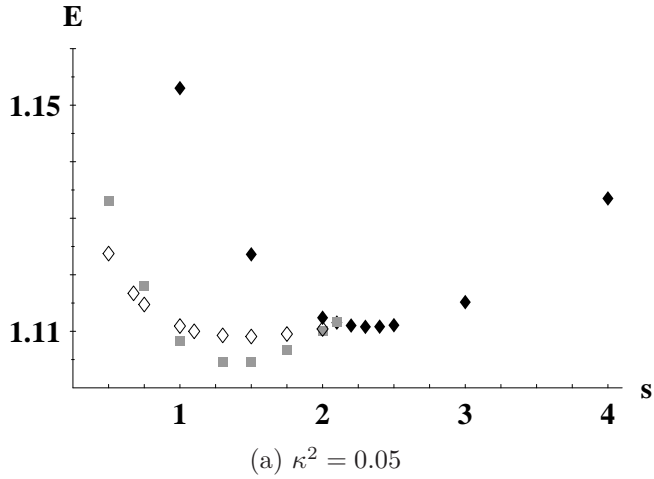}}
\hspace{1cm}
\break
\subfloat[$\kappa^2=0.25$]{
\label{s123b} 
\includegraphics[angle=0,scale=1,width=0.60\textwidth]{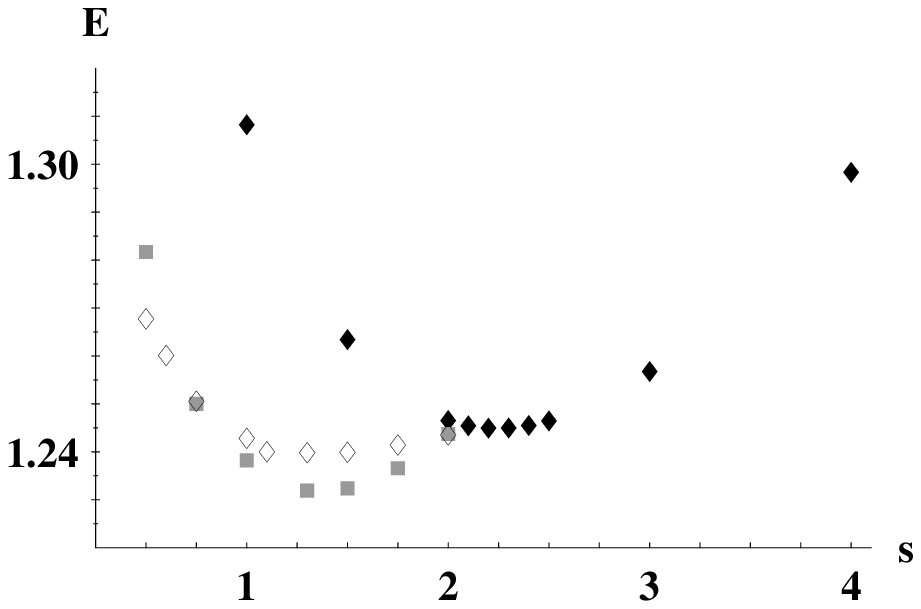}}
\hspace{1cm}
\break
\subfloat[$\kappa^2=1$]{
\label{s123c} 
\includegraphics[angle=0,scale=1,width=0.60\textwidth]{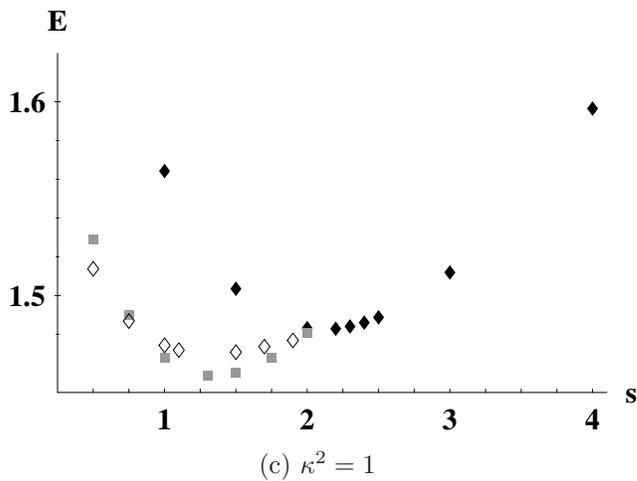}}
\caption{Total energies (divided by $4 \pi B$) of the charge--one ({\scriptsize $\fulldiamond$}) charge--two 
({\scriptsize \textcolor{Gray}{$\fullsquare$}})
and charge--three ({\scriptsize $\opendiamond$}) Skyrmions 
 as a function of the parameter $s$ for various $\kappa$ values.
Each of the energy graphs attains a minimal value at some $s$. At $s \approx 2$ the 
energy--per--topological--charge of the charge--two and charge--three solutions reaches
the charge--one energy (from below), and stable solutions are no longer observed.}
\label{fig:s123}
\end{figure}
\end{center}

Interestingly, the Skyrmion energy as function of $s$ is not monotonic,
but acquires a minimum at $s \approx 2.2$, as is shown in figure \ref{fig:s123}. 

\subsection{Charge--two Skyrmions}
\begin{figure}[htp!]
\includegraphics[angle=0,scale=1,width=1.1\textwidth]{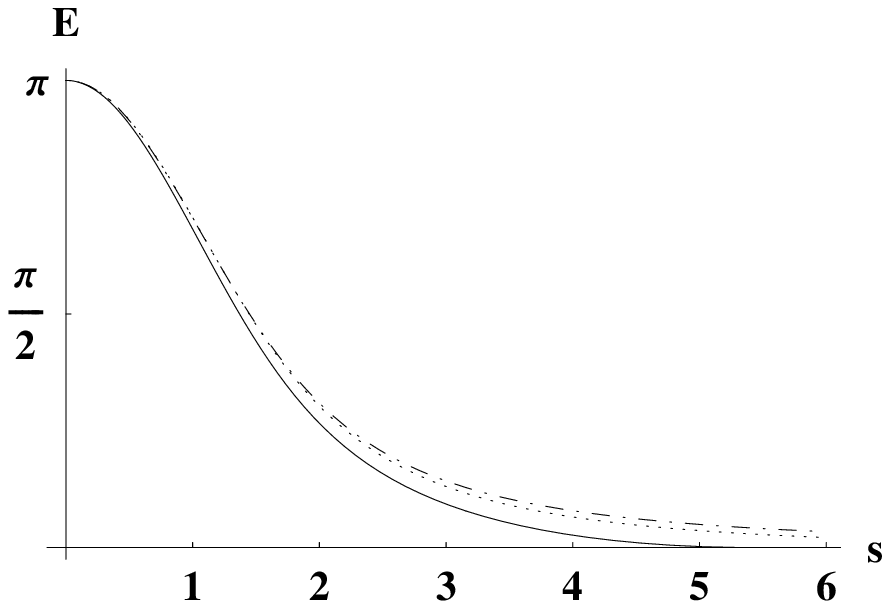}
\caption{Profile functions of the $B=2$ Skyrmion
for $s=0.5$ ($\full$), $s=1$ ($\dotted$) and $s=2$ ($\chain$). $\kappa$ is fixed at $\kappa^2=0.25$.}
\label{profB2}
\end{figure}

Stable solutions were also found 
to exist in the $B=2$ sector, but only for $s < 2$.
They are rotationally--symmetric and ring--like, corresponding to two charge--one
Skyrmions on top of each other. Figure \ref{profB2} shows the profile functions of
the stable solutions for different values of $s$ and  $\kappa^2=0.25$. 
\par
\begin{figure}
\subfloat[$s=1.5$]{
\label{contourB2a} 
\includegraphics[angle=0,scale=1,width=0.47\textwidth]{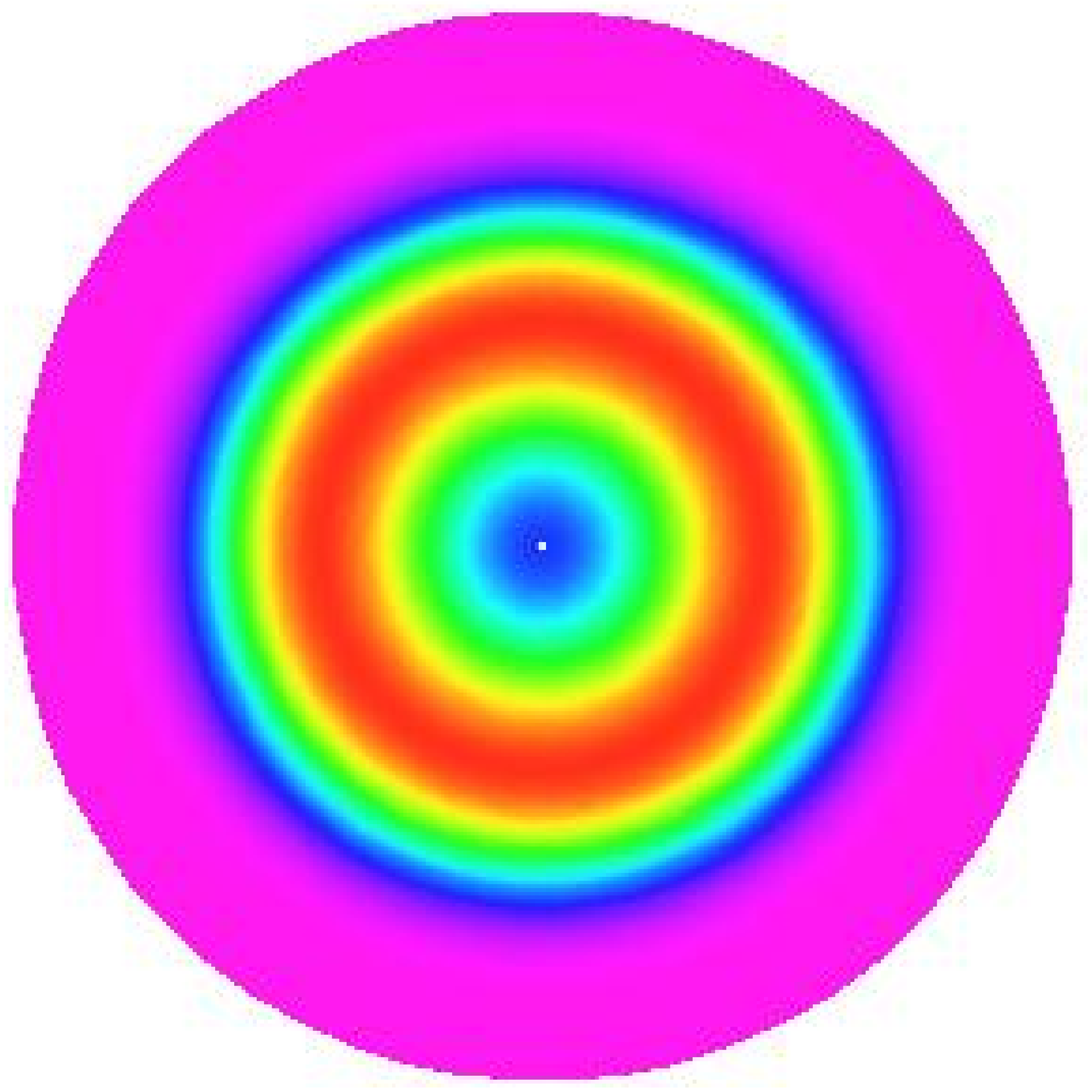}}
\subfloat[$s=2.6$]{
\label{contourB2b} 
\includegraphics[angle=0,scale=1,width=0.47\textwidth]{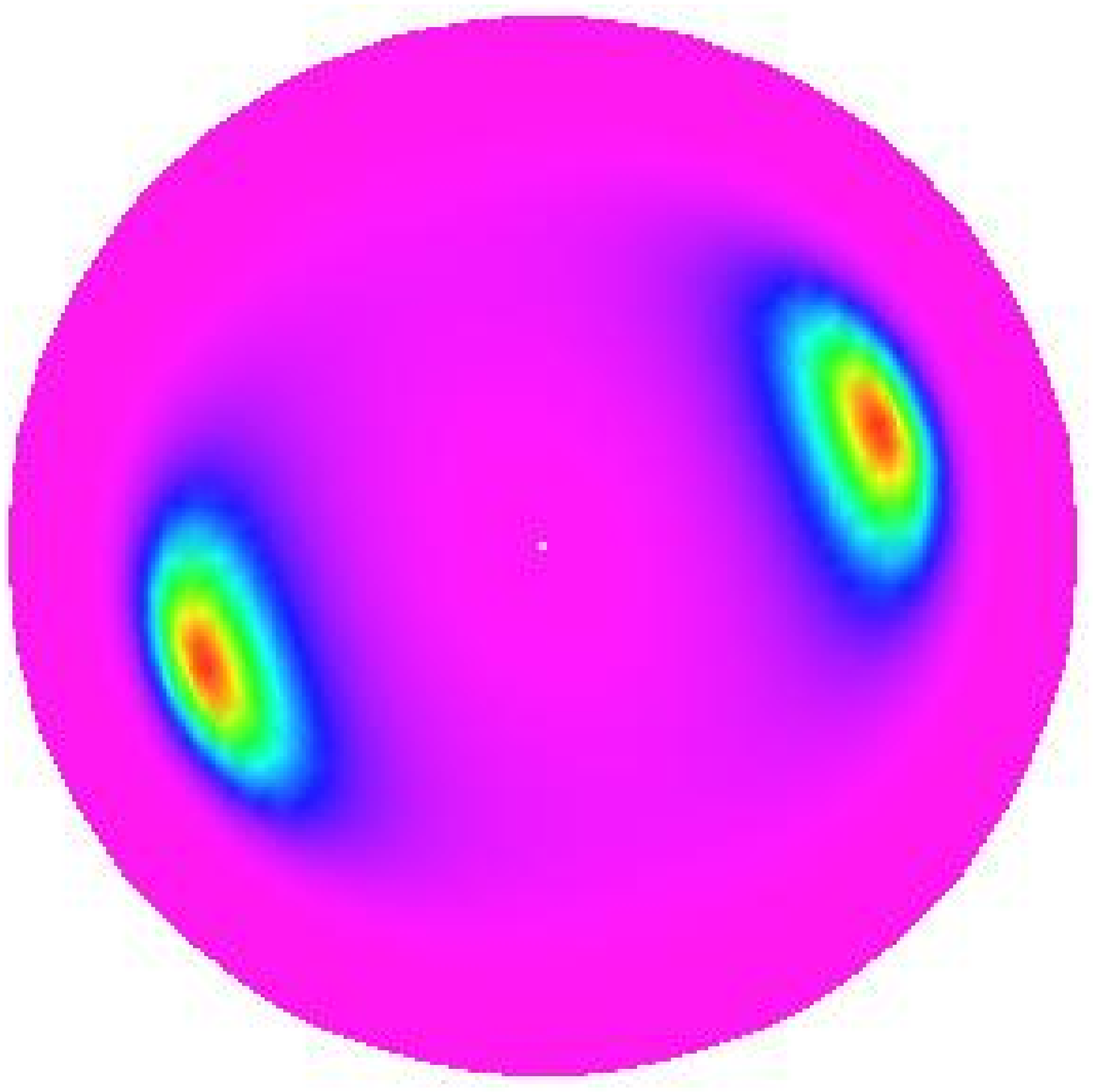}}
\caption{Contour plots of the energy distributions (ranging from violet -- low density 
to red -- high density) of the $B=2$ Skyrmion for $\kappa^2=1$. 
The energy is plotted as function 
of the angle $\theta$ and of the scaled radial coordinate $\tilde{r}$
(eq. (\ref{eq:truncR})). 
In the $s<2$ regime, ring--like rotationally--symmetric configurations exist,
corresponding to two charge--one Skyrmions on top of each other (left), whereas 
for $s>2$, the charge--two Skyrmion splits into two one--charge Skyrmions drifting infinitely apart (right).}
\label{contourB2}
\end{figure}

As in the $B=1$ case, the energy of the charge-two Skyrmion as function of $s$ is non--monotonic 
and has a minimum in $s$ around $s = 1.3$. 
As shown in figure \ref{fig:s123}, at $s \approx 2$ the energy of the ring--like configuration reaches the value of 
twice the energy of the charge--one Skyrmion and stable configurations cease to exist.
At this point, the Skyrmion breaks apart into 
its constituent charge--one Skyrmions, which in turn
start drifting away from each other. 
Contour plots of the energy distribution of the $B=2$ Skyrmion
are shown in figure \ref{contourB2} for $\kappa^2=1$ and for two $s$ values.
While for $s=1.5$ a ring--like stable configuration exists (figure \ref{contourB2a}),
for $s=2.6$ the Skyrmion breaks apart. The latter case is shown in figure \ref{contourB2b} which
was taken while the distance between the individual Skyrmions kept growing.
\par
These results are in accord
with corresponding results from previously known studies of both the `old' ($s=1$) model
in which ring--like configurations have been observed \cite{Old1,Old2}, and the holomorphic model 
for which no stable solutions have been found \cite{Holo1,Holo2}.

\par
We have also observed  rotationally--symmetric locally stable solutions 
in the large $s$ regime, including the `holomorphic' $s=4$ case.
The global minimum in this regime corresponds to 
two infinitely separated charge--one Skyrmions. 
The total energy of the rotationally symmetric solutions was found to be
larger than twice the energy of a charge--one Skyrmion, indicating
that the split Skyrmion is
an energetically more favourable configuration.
We discuss this issue in more detail in the section \ref{sec:res}.

\subsection{Charge--three and higher--charge Skyrmions}
As with the $B=2$ Skyrmion, the existence of stable charge--three
Skyrmions was also found to be $s$ dependent.
For any tested value of $\kappa$ in the range 
$0.01 \leq \kappa^2 \leq 1$, we have found that above $s \approx 2$, no stable charge--three solutions exist;
in this region the Skyrmion breaks apart into individual Skyrmions
drifting further and further away from each other. 
\par
\begin{center}
\begin{figure}[htp!]
\subfloat[$s=0.5$]{
\label{fixedKappaB3a} 
\includegraphics[angle=0,scale=1,width=0.29\textwidth]{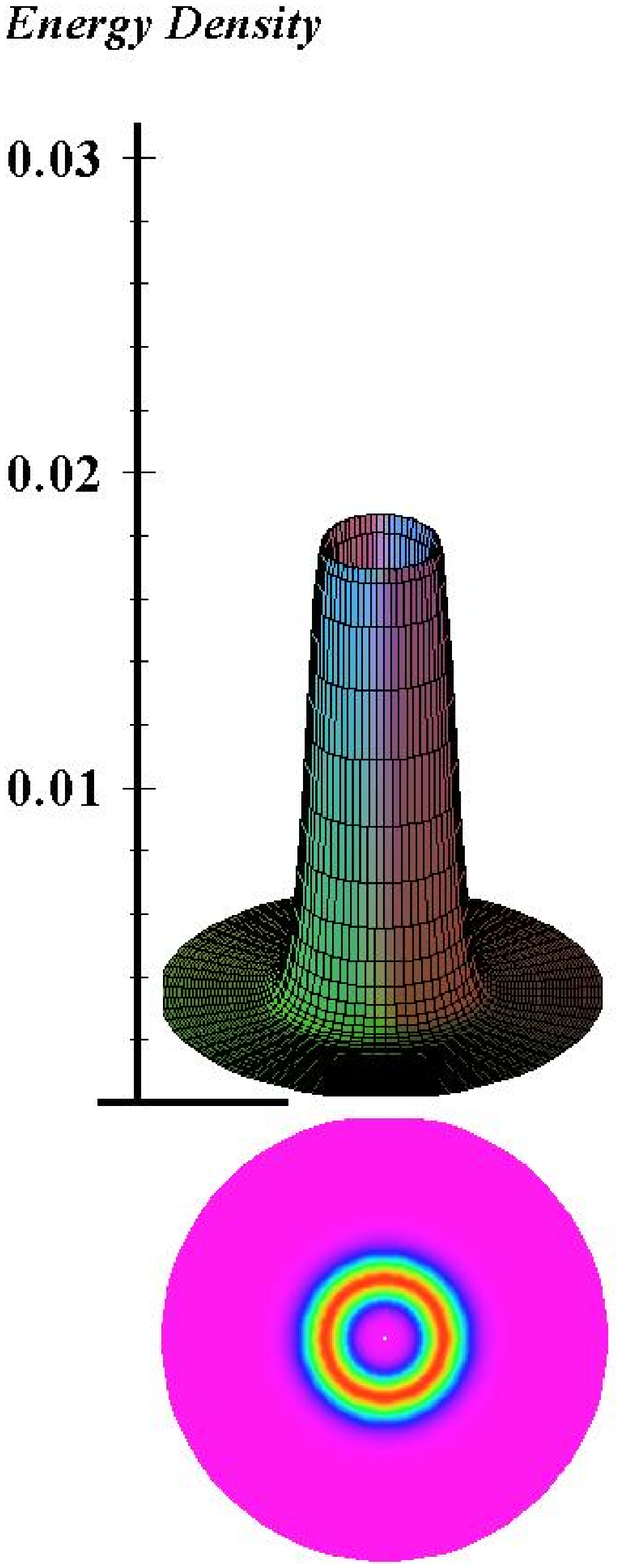}}
\subfloat[$s=0.75$]{
\label{fixedKappaB3b} 
\includegraphics[angle=0,scale=1,width=0.22\textwidth]{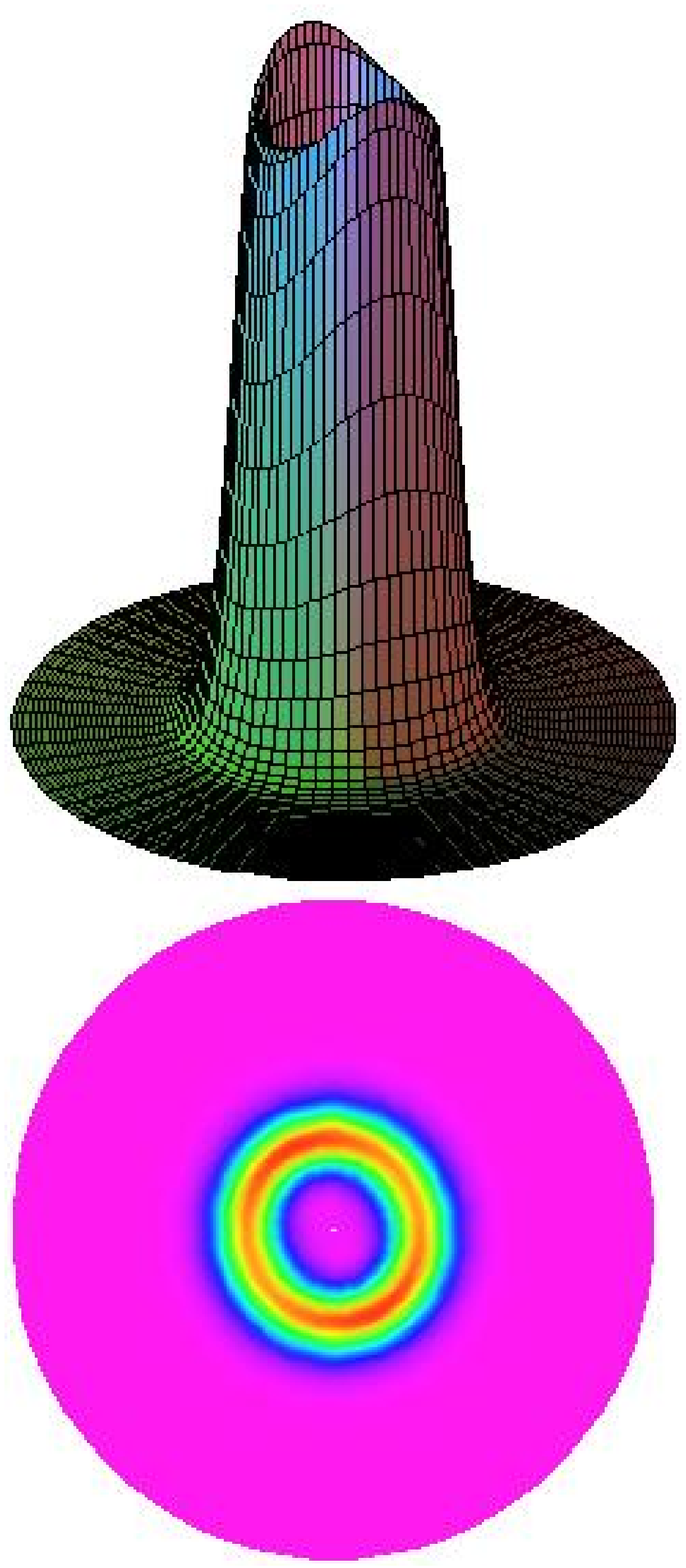}}
\subfloat[$s=1$]{
\label{fixedKappaB3c} 
\includegraphics[angle=0,scale=1,width=0.22\textwidth]{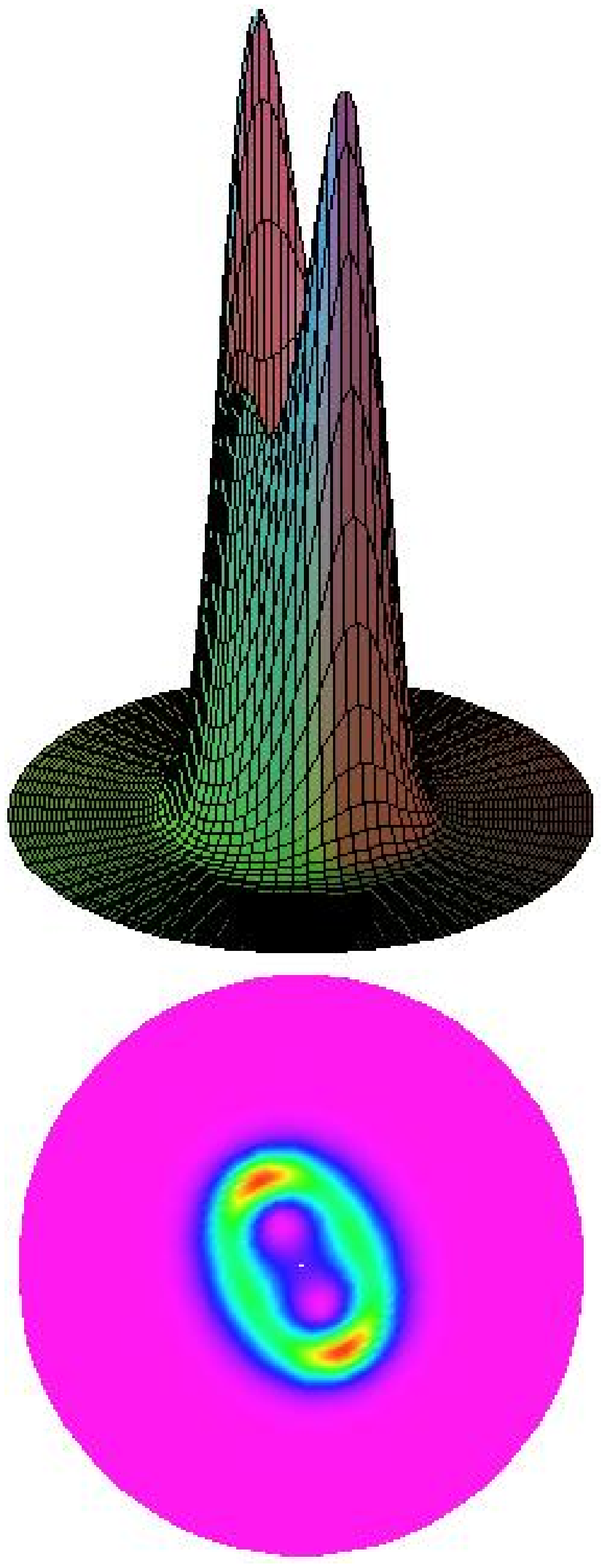}}
\subfloat[$s=3$]{
\label{fixedKappaB3d} 
\includegraphics[angle=0,scale=1,width=0.22\textwidth]{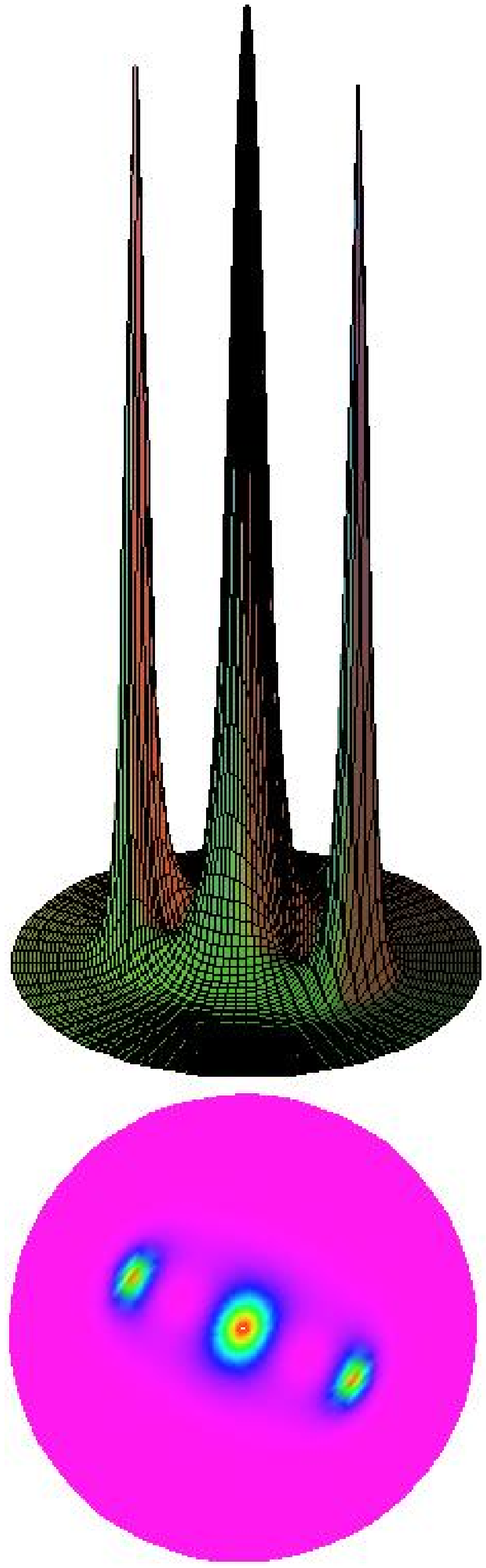}}
\caption{Energy densities and corresponding contour plots (ranging from violet -- low density 
to red -- high density) of the $B=3$ Skyrmion for fixed $\kappa$ ($\kappa^2=0.01$)
and varying $s$. 
The energy is plotted as function 
of the angle $\theta$ and of the scaled radial coordinate $\tilde{r}$
(eq. (\ref{eq:truncR})). 
In the $s=0.5$ case, the minimal energy configuration 
is rotationally symmetric, corresponding the three one--Skyrmions on top of each other.
For $s=0.75$ and $s=1$ the solutions exhibit only $\mathbb{Z}(2)$ symmetry,
corresponding to partially--overlapping one--Skyrmions. For $s=3$ no stable solution exists
and the individual Skyrmions are drifting apart. }
\label{fixedKappaB3}
\end{figure}
\end{center}

In the $s<2$ region, where stable solutions exist, 
the energy distribution of the charge--three Skyrmion
was found to be highly dependent on both $s$ and $\kappa$.
Keeping $\kappa$ fixed and varying $s$, we have found that in the small $s$ regime,
ring--like rotationally--symmetric configurations exist. 
Increasing the value of $s$ resulted in stable minimal energy configurations
with only $\mathbb{Z}(2)$ symmetry, corresponding to three partially--overlapping charge--one Skyrmions in a row,
as shown in figures \ref{fixedKappaB3b} and \ref{fixedKappaB3c}. 
The energy of the charge-three
Skyrmion also has a minimum in $s$, at around $s = 1.5$ (As shown in figure \ref{fig:s123}). At $s \approx 2$ 
the energy of the Skyrmion becomes larger than three times the
energy of a charge--one Skyrmion and stable configurations are no longer obtainable.
This is illustrated in figure \ref{fixedKappaB3} which shows 
contour plots of the energy distribution of the $B=3$ Skyrmion
for different values of $s$ and fixed $\kappa$. For $s=0.5$ (figure \ref{fixedKappaB3a}), the solution
is rotationally symmetric and 
for $s=0.75$ and $s=1$ (figures \ref{fixedKappaB3b} and \ref{fixedKappaB3c} respectively)
the rotational symmetry 
of the solution is broken and only $\mathbb{Z}(2)$ symmetry remains. At $s=3$, no stable solution exists. 
The latter case is shown in figure \ref{fixedKappaB3d} which
was taken while the distance between the individual Skyrmions
kept growing.
\par
\begin{figure}[htp!]
\subfloat[$\kappa^2=0.01$]{
\label{fixedSB3a} 
\includegraphics[angle=0,scale=1,width=0.3\textwidth]{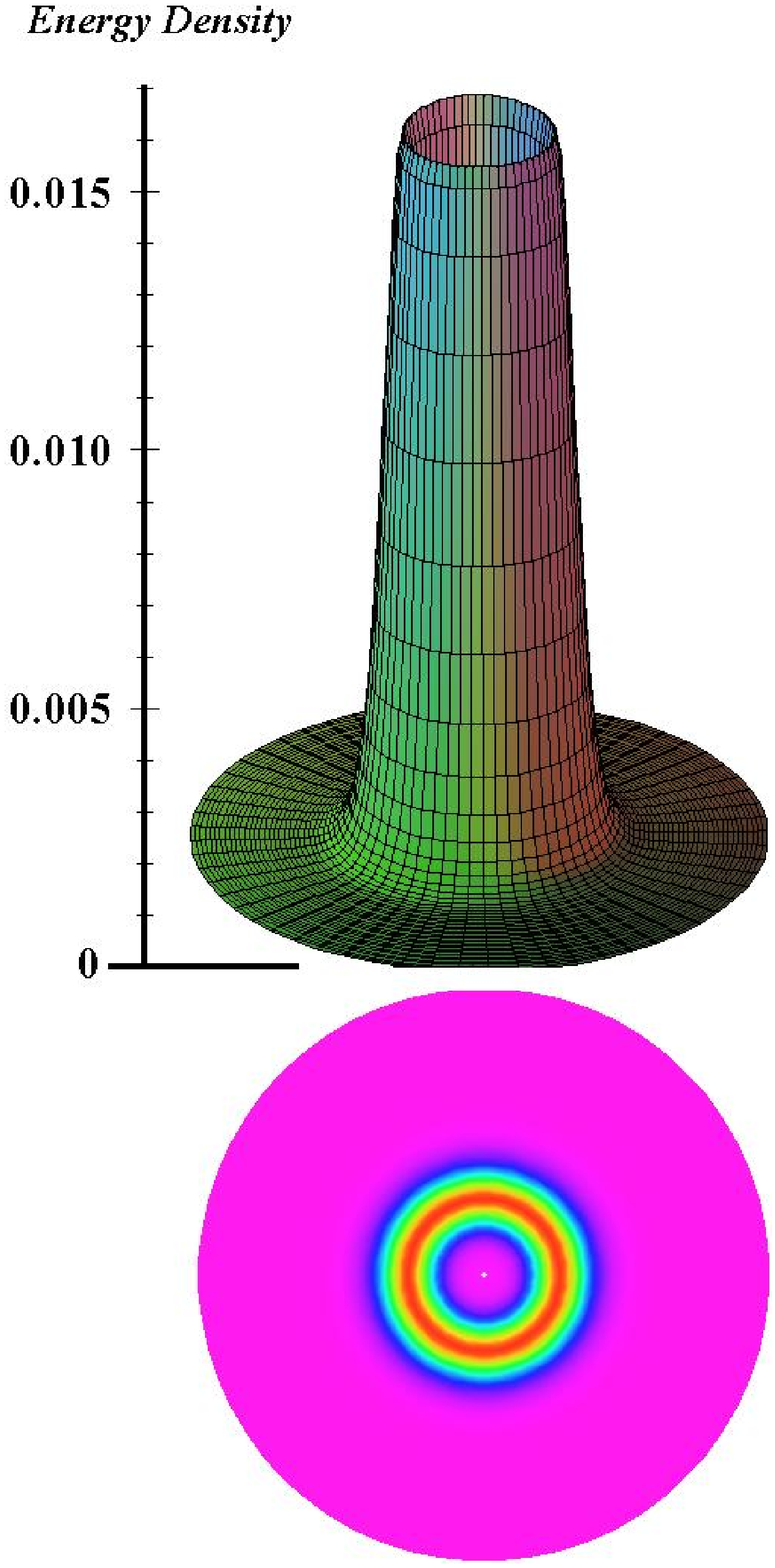}}
\subfloat[$\kappa^2=0.1$]{
\label{fixedSB3b} 
\includegraphics[angle=0,scale=1,width=0.23\textwidth]{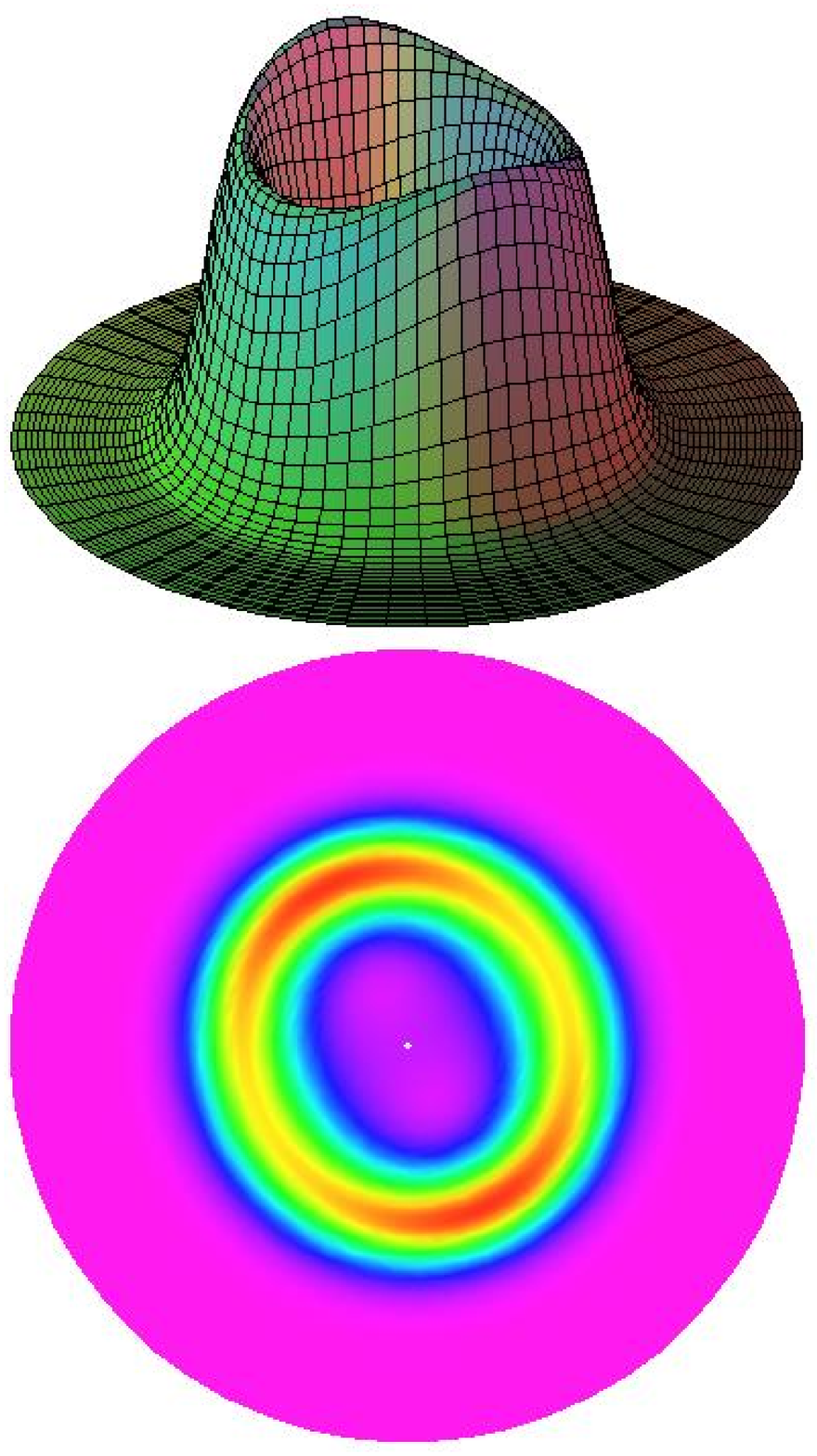}}
\subfloat[$\kappa^2=0.25$]{
\label{fixedSB3c} 
\includegraphics[angle=0,scale=1,width=0.23\textwidth]{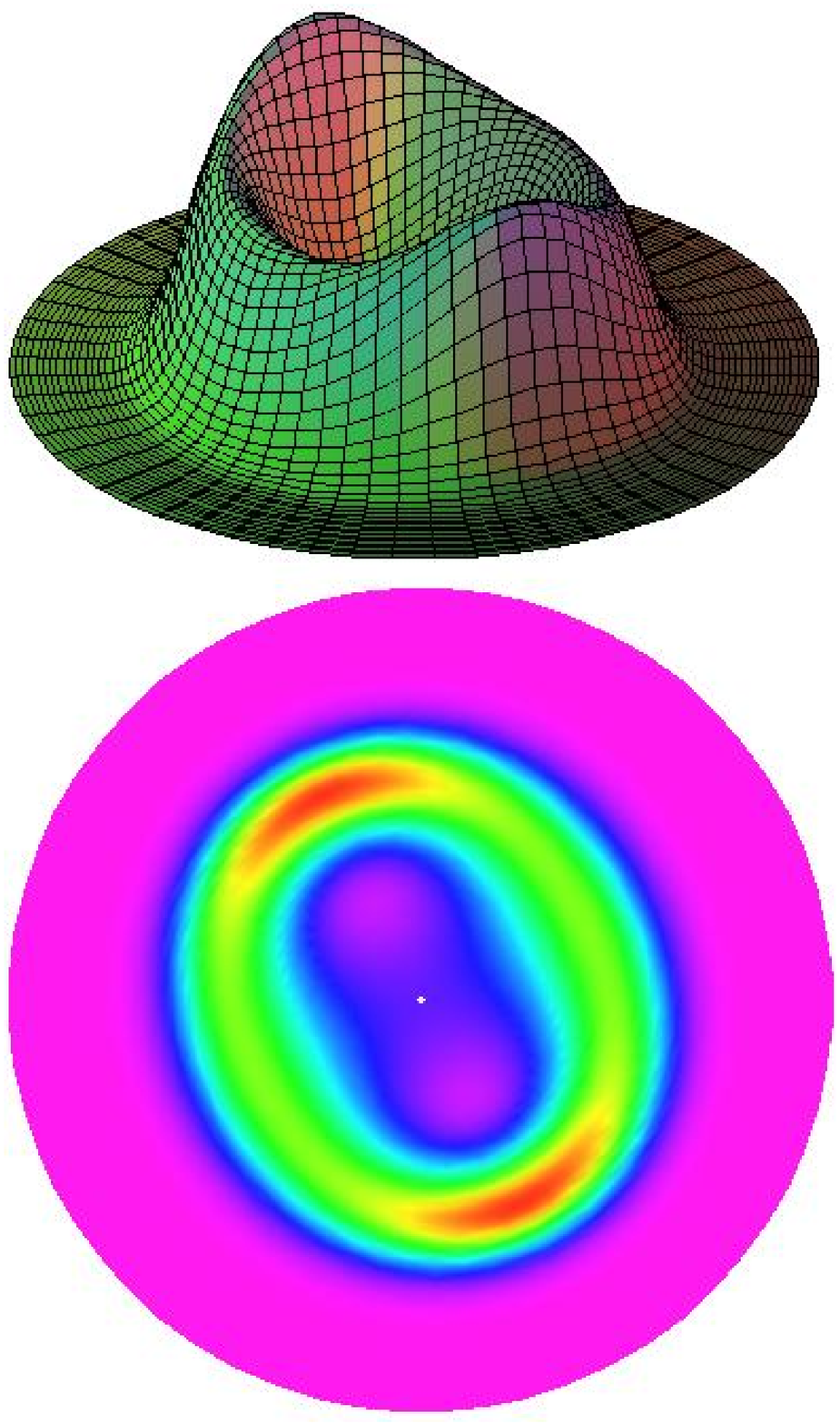}}
\subfloat[$\kappa^2=1$]{
\label{fixedSB3d} 
\includegraphics[angle=0,scale=1,width=0.23\textwidth]{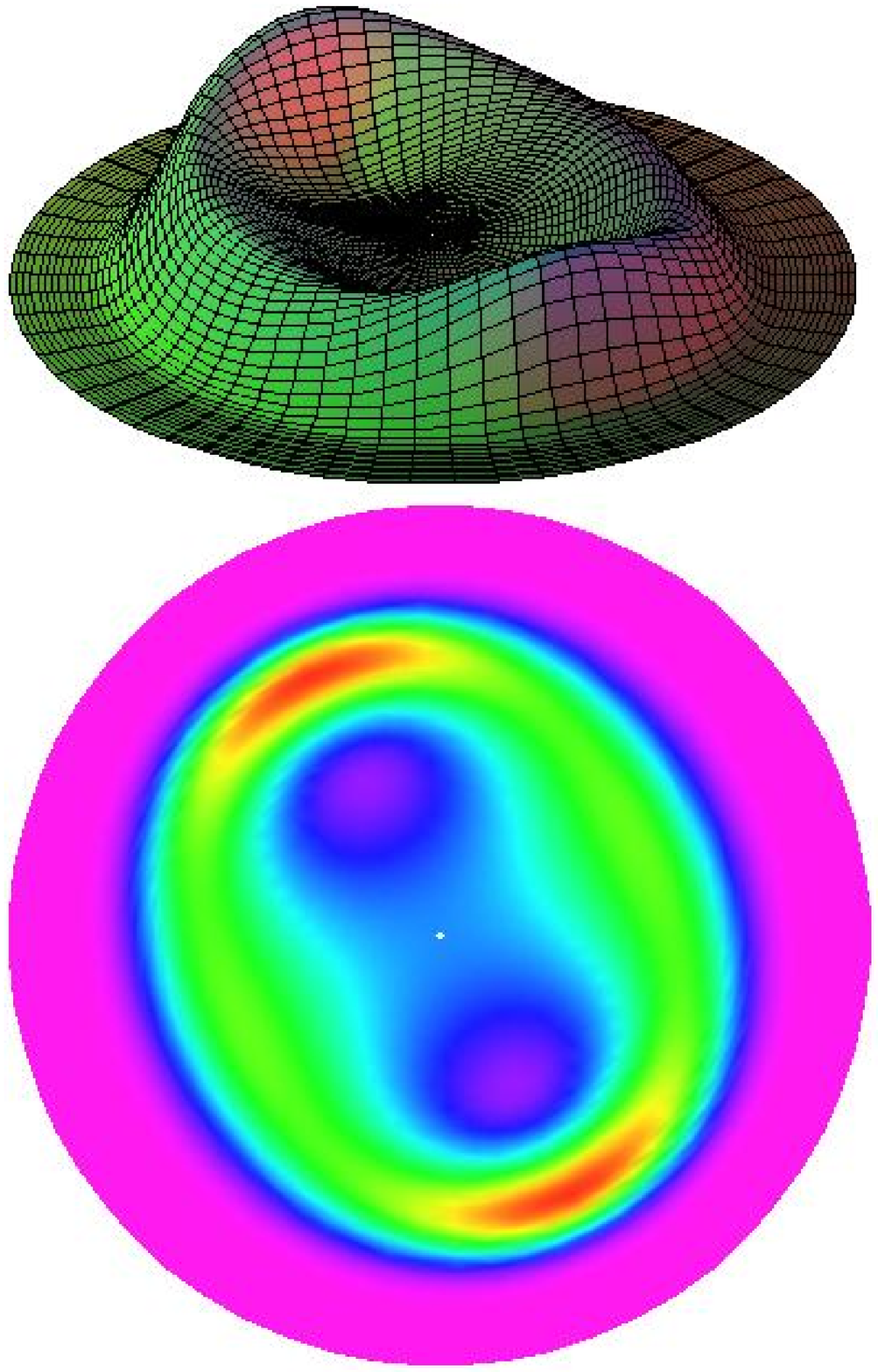}}
\caption{Energy densities and corresponding contour plots (ranging from violet -- low density 
to red -- high density) of the $B=3$ Skyrmion for fixed $s$ ($s=0.5$)
and varying $\kappa$. At low $\kappa$, the minimal energy configuration is rotationally
symmetric. As $\kappa$ is increased, breaking of rotational symmetry appears,
and only $\mathbb{Z}(2)$ symmetry remains.}
\label{fixedSB3}
\end{figure}
\begin{center}
\begin{figure}[hbp!]
\includegraphics[angle=0,scale=1,width=1.1\textwidth]{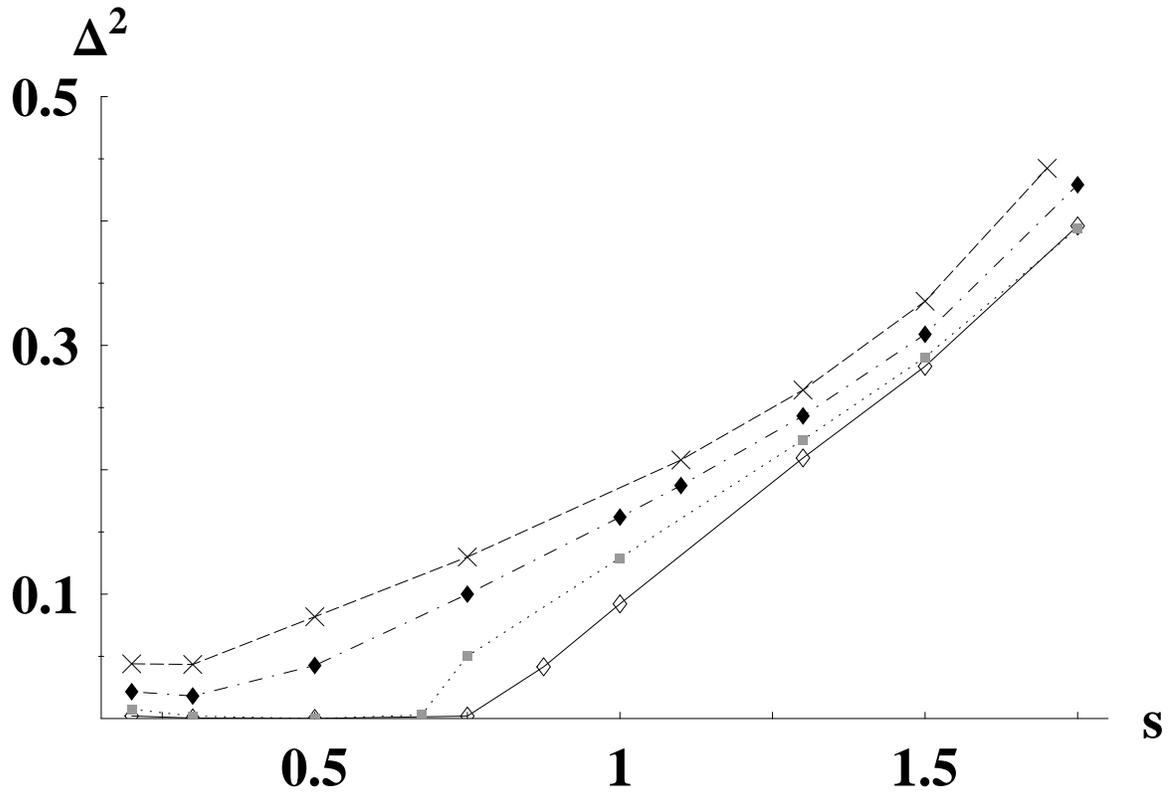}
\caption{ Symmetry--breaking measure $\Delta^2$, 
as a function of $s$ for various $\kappa$ values in the charge--three sector:
$\kappa^2=0.01$ ({\scriptsize $\opendiamond$}), $\kappa^2=0.05$ ({\scriptsize \textcolor{Gray}{$\fullsquare$}}),
$\kappa^2=0.25$ ({\scriptsize $\fulldiamond$}), and
$\kappa^2=1$ ($\times$). Breaking of rotational symmetry becomes more and more apparent
as $s$ and $\kappa$ increase. The lines are to guide the eye.}
\label{Delta2}
\end{figure}
\end{center}

The dependence of the Skyrmion solutions 
on the value of $\kappa$ with fixed $s$
has also been studied. Qualitatively, for small $\kappa$ the minimal energy configurations are rotationally--symmetric.
Increasing its value results
in an increasingly larger rotational symmetry breaking. This is illustrated in
figure \ref{fixedSB3}. 
\newpage
To better appreciate the effects of the parameters of the model
on the obtained solutions, we 
introduce the following measure for the breaking of rotational symmetry:
\bea
\Delta^2= \int\left( \frac1{2 B} \int \mathcal{B}(r,\theta) r \rmd r \right)^2 \frac{d \theta}{2 \pi}   -1 \;,
\eea
where $\mathcal{B}(r,\theta)$ is the charge density of the Skyrmion with rotationally--symmetric 
configurations corresponding to $\Delta^2=0$.
In terms of this measure, the results presented above may be quantitatively characterized.
In figure \ref{Delta2}, $\Delta^2$ 
is plotted as function of the potential parameter $s$  for various $\kappa$ values,
illustrating the effects of these parameters on the breaking of
rotational symmetry of the charge--three solutions.  

\begin{figure}[hbp!]
\subfloat[$B=4$, $s=0.9$]{
\label{Contour45a}
\includegraphics[angle=0,scale=1,width=0.47\textwidth]{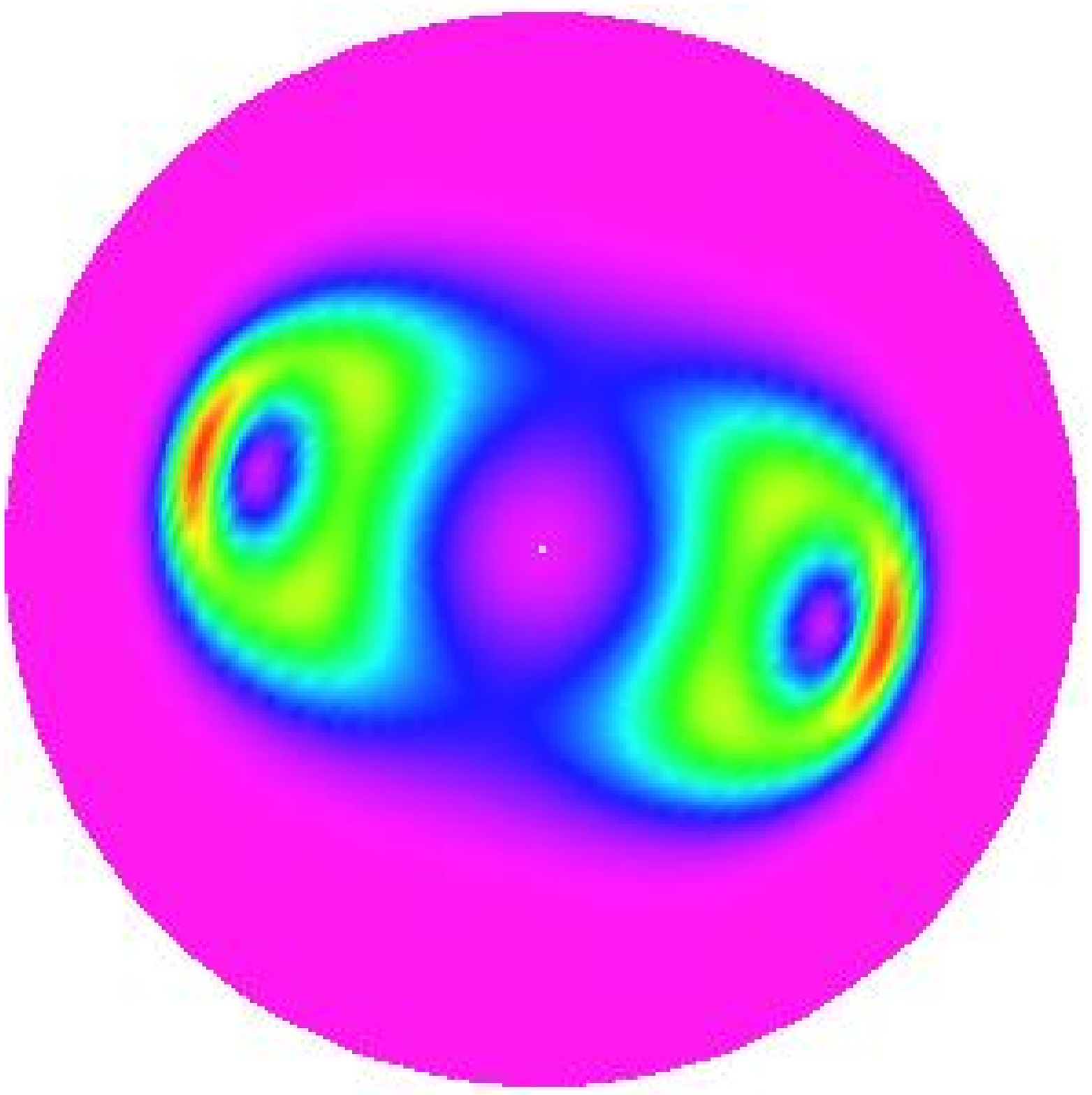}}
\centering
\subfloat[$B=5$, $s=0.9$]{
\label{Contour45b}
\includegraphics[angle=0,scale=1,width=0.47\textwidth]{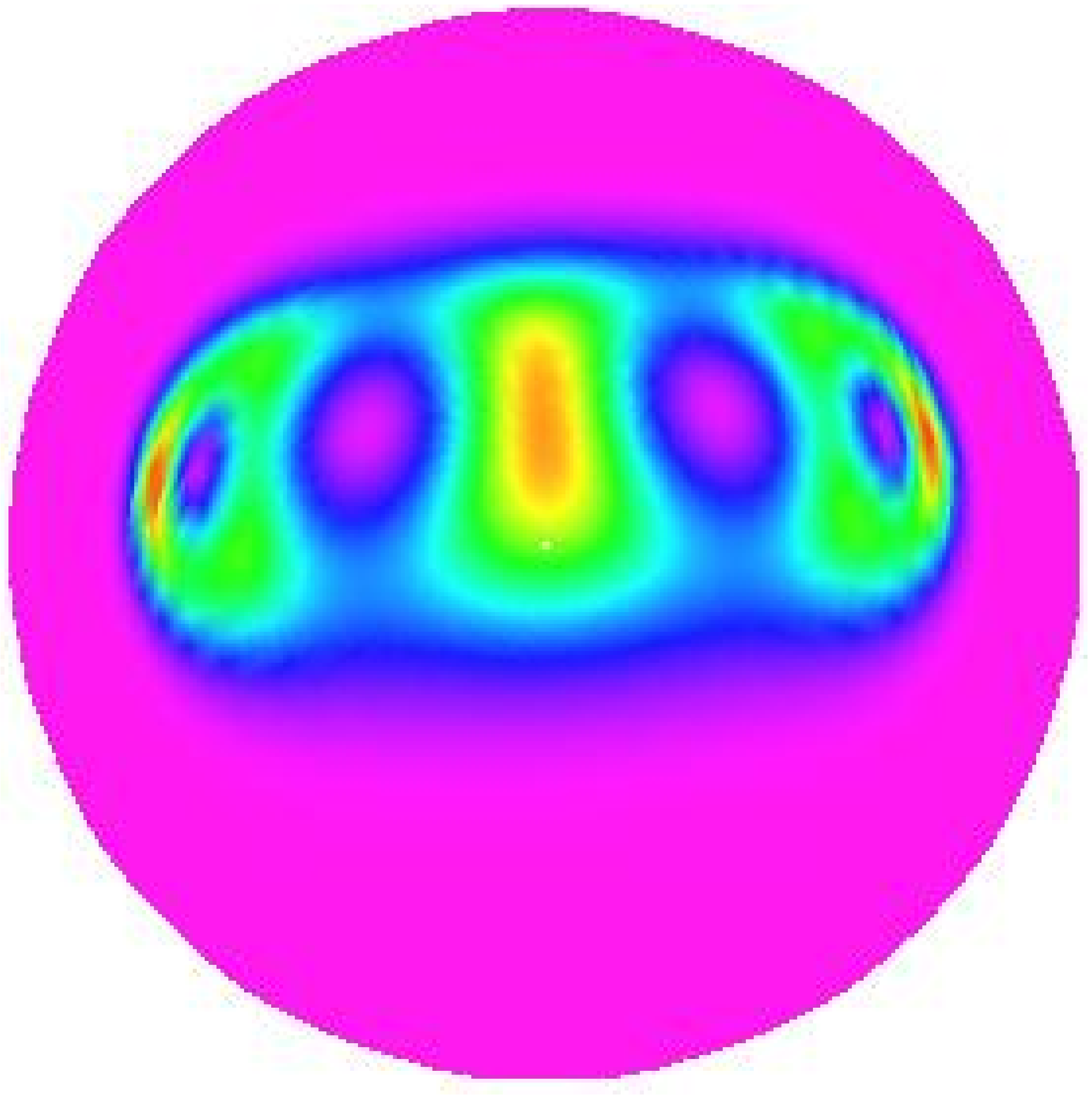}}
\hspace{1cm}
\subfloat[$B=4$, $s=4$]{
\label{Contour45c}
\includegraphics[angle=0,scale=1,width=0.47\textwidth]{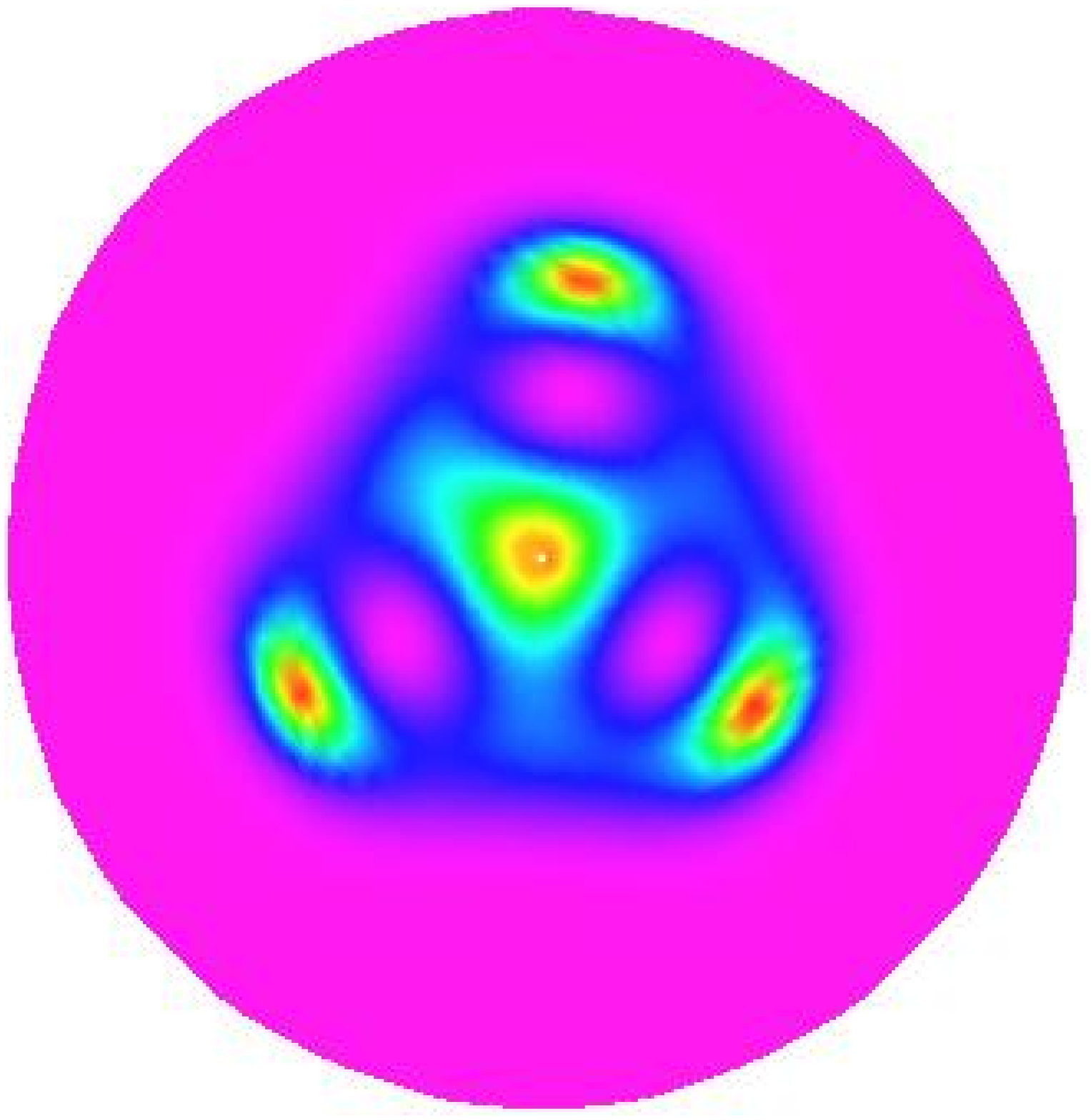}}
\subfloat[$B=5$, $s=4$]{
\label{Contour45d}
\includegraphics[angle=0,scale=1,width=0.47\textwidth]{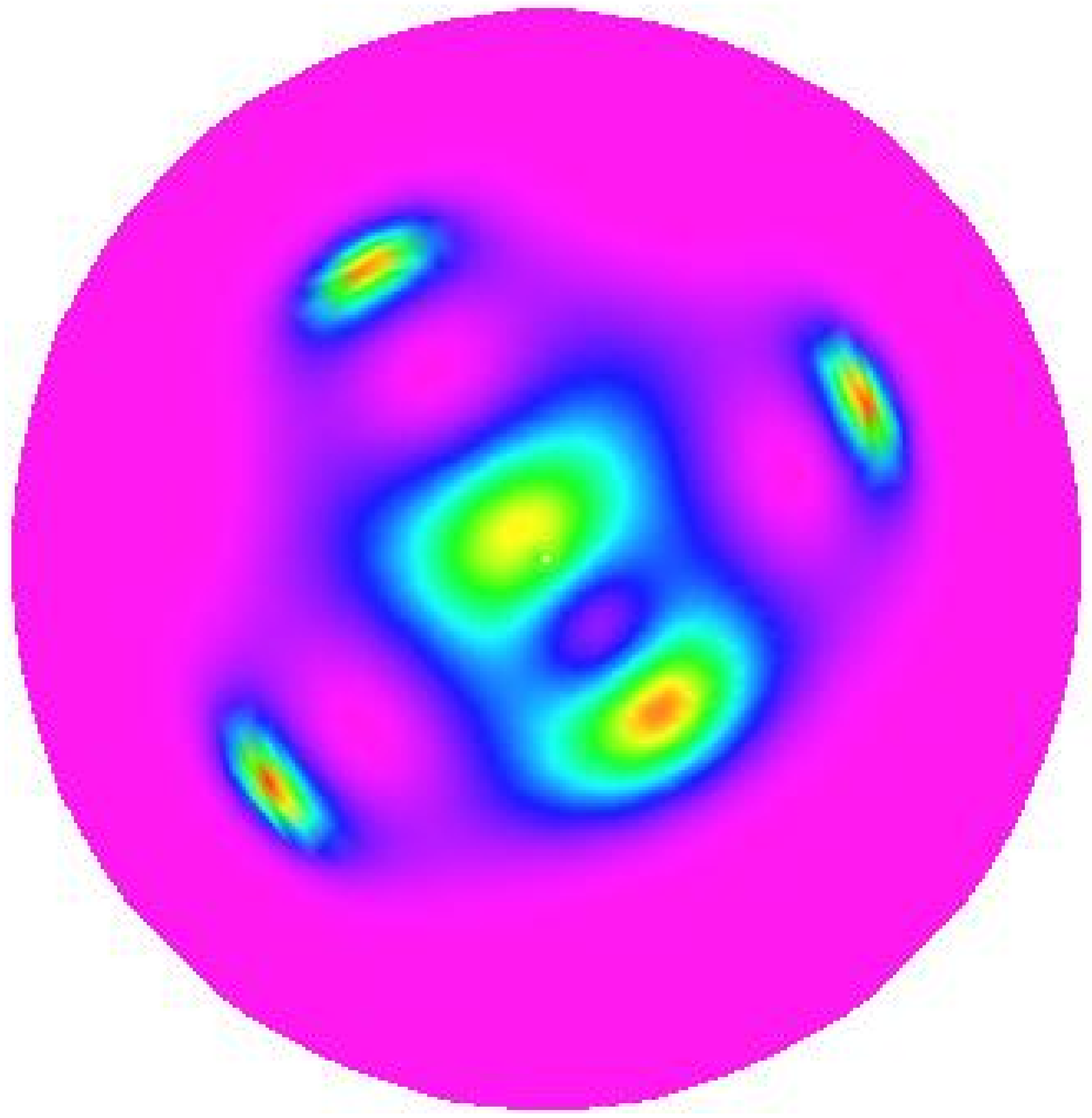}}
\caption{Contour plots of the energy distributions (ranging from violet -- low density 
to red -- high density) of the $B=4$ and $B=5$ Skyrmions
for $s=0.9$ and $s=4$ ($\kappa^2=0.1$). 
In the lower $s$ region stable solutions exist; the upper figures show a $B=4$ Skyrmion 
in a bound state of two charge--two Skyrmions (left), and a $B=5$ Skyrmion 
in a two--one--two configuration. For values of $s$ higher than $2$, 
the multi--Skyrmions split into individual one--Skyrmions
constantly drifting apart (lower figures).}
\label{Contour45}
\end{figure}

The $B=4$ and $B=5$ Skyrmion solutions behave similarly
to the $B=3$ solutions, although (due to much larger computation time requirements)
here the parameter space was not scanned as densely as with the lower charges. 
An illustrative example is given in figure \ref{Contour45},
which shows the stable solutions that have been obtained in the $s=0.9$ case 
and the splitting of these Skyrmions
into their constituents in the $s=4$ case.
\clearpage
\section{\label{sec:res} Summary and conclusion}
We have studied the static solutions of the baby Skyrme model
for the one--parametric family of potentials $U=\mu^2(1-\phi_3)^s$.
While Skyrmions with charge one were found 
to be stable and rotationally--symmetric for all tested values of $\kappa$ and $s$,
the symmetries  and stability of the minimal energy configurations of Skyrmions
with higher charge were found to be strongly 
dependent on the parameter of the potential. 
\par
The charge--two Skyrmion was found to be ring--like and rotationally--symmetric
but only for $s < 2$, implying that in this regime, there exists
a strong attraction between the individual Skyrmions.
Above $s \approx 2$, the two--Skyrmion breaks apart into one--Skyrmions which repel each other and as
a result drift infinitely apart.
\par
For the $B>2$ Skyrmions, small values of $s$ yielded in general
rotationally symmetric configurations; 
implying that the attraction between the individual Skyrmions is strong.
Increasing the value of $s$ resulted in stable configurations with broken rotational symmetry.
This type of solutions implies only a moderate attraction between the
constituent Skyrmions, as this configuration  may be viewed as three partially--overlapping 
charge--one Skyrmions in a row.
As in the $B=2$ case, for $s>2$ no stable configurations exist.
This is a consequence of a very strong repulsion between 
the Skyrmions which drives them infinitely apart from each other.\par
Altogether, it seems that $s$ may be viewed
as a `control' parameter for the strength of the attraction 
(small $s$ values) or repulsion (large $s$ values)
between individual Skyrmions. 
\par
We also note here the locally stable solutions
that have been obtained 
in the $B=2$ sector. We particularly note the locally stable two--Skyrmion 
solution that has been found in the `holomorphic' $s=4$ case.
This result stands in contradiction with previously known results \cite{Holo3}
where it has been found that no stable solutions exist in this case.
This discrepancy is not fully understood and might be due to the
differences between the method used in \cite{Holo3} which
is based on the time evolution of a system of two well separated Skyrmions 
sent towards one another, and the current one, in which multisolitons are obtained 
in a rather different approach, namely, by minimizing the energy functional of the system.
Bearing in mind that the initial configuration in our minimization scheme 
is rotationally symmetric, it is possible
that the minimization procedure ends up at a local rotationally--symmetric minimum,
provided such a local minimum exists.
\par
We verified that the results of our minimization procedure 
are robust with respect to changes in the grid size and in the initial configuration.
They were also verified using the ``simulated annealing''.
In most cases (with some notable exceptions as discussed above),
our results were also found to agree with
previously known results whenever a comparison was available.  
\ack
We thank Wojtek Zakrzewski for useful discussions. 
This work was supported in part by a grant from the Israel Science
Foundation administered by the Israel Academy of Sciences and Humanities.

\end{document}